\documentclass{elsart}

\usepackage{epsfig}
\usepackage{amssymb,amsmath}
\makeatother
\begin{document}
%\begin{frontmatter}
%\topmatter
\title
{A Collective Motion Algorithm for Tracking Time-Dependent Boundaries}
%\endtitle
\author{Ioana Triandaf and Ira B. Schwartz}
\address{
Naval Research Laboratory, Code 6792, Plasma Physics Division, Washington, DC 20375 }
%\date{\today}
\begin{abstract}
We present a numerical method that allows a formation of communicating agents
to target the boundary of a time dependent concentration by
following a time dependent concentration gradient.
The algorithm motivated by \cite{MB, BKM}, allows
the agents to move in space in a non-stationary environment.
Our method is applicable to finding the boundary of any regular
surface and may be of interest for
for studying motions of swarms in biology, as well as to engineering
applications where boundary detection is an issue. 
\end{abstract}

\maketitle
%\endtopmatter 
%\end{frontmatter}

\section{Introduction}

We propose and study a method for a multi-agent system of autonomous vehicles to perform
the exploration of a non-stationary environment. The goal 
of this exploration is to reach a designated target geometry, typically a
boundary, and to describe that boundary. The current paper relates to the 
general objective of studying motions of swarms, which we understand as a 
collection of autonomous entities which rely on local sensing and simple
behavior, interacting in a way that a more complex
behavior of the whole group emerges from local interactions \cite{CP}. 
This type of behavior is well-known in natural phenomena such as schools of
fish  \cite{CP} and colonies of insects \cite{BDT}. One can find in the
literature various other methods of simulating motions of
distributed systems inspired by biology \cite{BDT}, \cite{PEK},
and by economics-based concepts \cite{SD}. 
The applications of such motions are quite broad since substituting agents for
humans is desirable in many areas such as de-mining operations \cite{CASS},
mapping and exploration \cite{BMFST}, navigation control \cite{DOK}, formation flying
\cite{OFL} and military planning \cite{CP}. 

The algorithm presented here models a group of agents to move on a given
surface or within a prescribed volume, find its boundary and track it, 
while communicating with each other and spreading simultaneously over the surface. The
environment in which the motion takes place 
varies in time. Such a method may provide insight 
to a better understanding of motions of swarms in biology \cite{BDT}.
The motion of swarms in military applications has been considered in
recent years \cite{CP}, since it offers an unmanned, dependable, flexible, 
self-organized type of network to achieve various military objectives, 
while replacing the burden of comprehensive communication, 
since the swarms interact primarily with their neighbors. 

The particular goal of identifying and tracking the boundary of a surface
which we address here, is analogous to the problem encountered in image
segmentation of tracing the boundary of an object.
Tracking contours as a function of time has been addressed  
in the pioneer work of Kass \cite{KWT}, in relation to localizing 
features in an image. Here we add more  complexity to the situation in \cite{KWT} 
since the motion of the agents starts far away from the boundary. Moreover, once the
boundary is located the contour of agents moves along that boundary 
rather than keeping a fixed location.

The idea of adapting image processing techniques to simulate motions of swarms
was proposed in \cite{BKM}, where the image segmentation technique of locating
a contour with a model known as a snake \cite{KWT}, or energy-minimizing curve, 
\cite{Sapiro} is adopted. We extent that analogy
by allowing the formation of agents to move on a time-dependent surface
or on a surface whose boundary travels in time.
In the current work we explore  the problem of simulating the motion of
agents which effectively locate and follow the boundary of a surface of interest,
assuming that the boundary is known to each agent and it is moving in time 
possibly along with the entire surface. The motion of agents takes place in
either 2D or 3D. 

The idea of the algorithm we present is to model the group of agents as a contour that
deforms towards the boundary of the object. Interactions between nearby agents
are assumed, the modeling of these interactions being analogous to having
forces acting between nearby particles as in an elastic band. An energy
functional results from this physical model in a continuum limit. 
The desired motion of the agents is obtained by minimizing this functional 
to which an appropriate goal dynamics term is added, so that its minimum is
obtained as the agents land on the proposed target. 

The algorithm assumes that each agent has the position of the boundary, has
information on nearby agents, and evaluates its position on the surface as well as
the local gradient of that surface. The movement of each agent can be modeled
by a partial differential equation for the velocity of each vehicle. 
By discretizing this equation we derive a numerical scheme for the motion of
each agent. Each equation contains a term accounting for movement 
along the steepest descent on the surface and a term corresponding to movement
parallel to the tangent to the boundary that is being tracked. 
Coordinated behavior is created by imposing local interaction rules among
agents. These terms also dictate the manner in which the formation will
converge towards the desired goal.

The paper is organized as follows: In the next section we present an algorithm 
for an individual agent to accomplish the desired task. 
Then we continue in section III with 
adding a sparsing term which connects the motions of nearby swarmers and
prevents collisions of agents in real applications.
Obtaining collective motion local communication rules is essential.
Typically these rules determine a general pattern of significance in
applications \cite{CP}. Consequently in section III we present a PDE model
that includes interaction between nearby swarmers. These interactions 
are based on modeling the swarm as the discretization of an elastic string
with elastic and bending forces acting between nearby particles.
In section IV we give a numerical
method based on the model in the previous section.  
We continue in section V with the corresponding algorithm in 3D, 
which means the motion now takes place on a surface defined in 3D rather then 2D. 
We end with a conclusion section.

\section{The Collective Motion Algorithm}

The simplest version of the algorithm consists in
designing motion terms that would lead a single vehicle to move towards a given
boundary and then move along that boundary. The component of the motion that determines 
the approach to a prescribed boundary consists in moving along the steepest
descent direction along a surface. To make the vehicle describe the boundary
we include a term that is parallel to 
the tangent to the boundary of the surface. The velocity of a vehicle is
thus calculated from these two components giving a system of two ordinary
differential equations. 

Let $C(x,y)$ denote the concentration function of the environment at the
agent's position $(x,y)$. 
By ``concentration'', we mean a function which describes any spatially varying 
variable, such as density, temperature, etc.
Let $P(x,y)=f(C(x,y))=(C-C_{0})^2$ be a function that achieves a
minimum at the boundary of the environmental concentration. Let $v=v(x,y)$
denote the position vector. The motion according to the rule $\frac {dv}{dt}=
-\nabla P$ results in a gradient descent toward a local minimum of the function
$P$, therefore towards the curve $C=C_{0}$. 
Also the rule  $\frac {dv}{dt}=\nabla ^{\perp}C$, (where $\nabla ^{\perp}C$
denotes the normal vector to the boundary of the surface) 
results in travel along the boundary
of the concentration surface $C$. Combining the two terms mentioned above
we arrive at the following model:

\begin{eqnarray}
\label{equation:Particle system}
\frac{dx}{dt} & = & -\partial_x P - \omega \frac {\partial _y C}{\|\nabla
  C\|}\nonumber \\
\label{rateequations}\\
\frac{dy}{dt} & = & -\partial_y P + \omega \frac {\partial _x C}{\|\nabla
  C\|}\nonumber \\ 
\nonumber
\end{eqnarray}
The parameter $\omega$ in Eq. ~\ref{equation:Particle system}  determines the speed of the vehicle in the direction of
the tangent to the boundary. 
Following \cite{BKM} as a test case, we consider as the concentration function: 
\begin{eqnarray}
\label{equation:Constraint}
C(x,y) & =  & tanh (F(x,y) - \frac {3}{4})\nonumber \\
\label{confunction}\\
F(x,y) & =  & \sum_{i=1}^{4} exp(\frac {-((x-x_i)^2+(y-y_i)^2)}{\sigma
  ^2}).\nonumber \\
\nonumber
\end{eqnarray}
Equation ~\ref{equation:Constraint}  is defined on a disk of radius 2 and $(x_i,y_i)=(1,0)$,
$(0,-\frac {1}{2}),(-\frac {3}{2},\frac {1}{2}),(\frac {3}{20},1)$ for
$i=1,2,3,4$ and $\sigma =1.0$.
$C_0$ is the concentration on the boundary, so it is obtained from Eq.
(~\ref{confunction} ) 
taking $x$ and $y$ on the boundary of the circle of radius 2.0 centered 
at the origin. In this way we are targeting a boundary that varies in space.

\begin{figure}
\centerline{\epsfig{figure=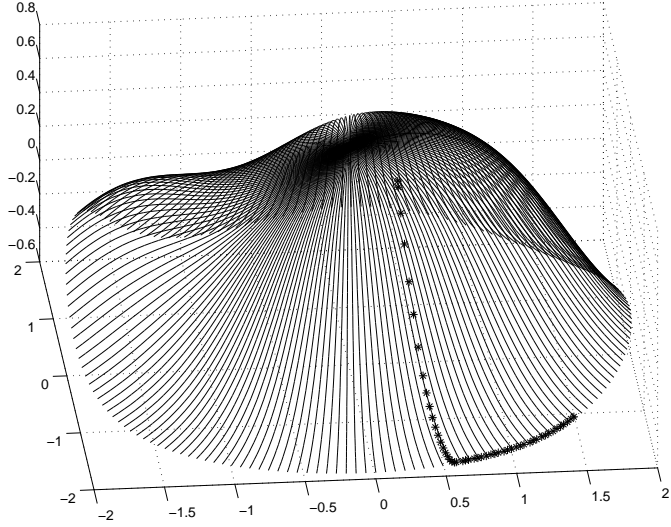, width=3.5in}}
\caption{\label{fig:singleagent} 
The trajectory of a single agent (represented by '*'s) launched on a
concentration surface,
at initial condition $x0=0.3$ and $y0=0.5$,
reaching the boundary and accurately following the boundary. Solid
lines denote the surface $C(x,y)$.}
\end{figure}

\begin{figure}
\centerline{\epsfig{figure=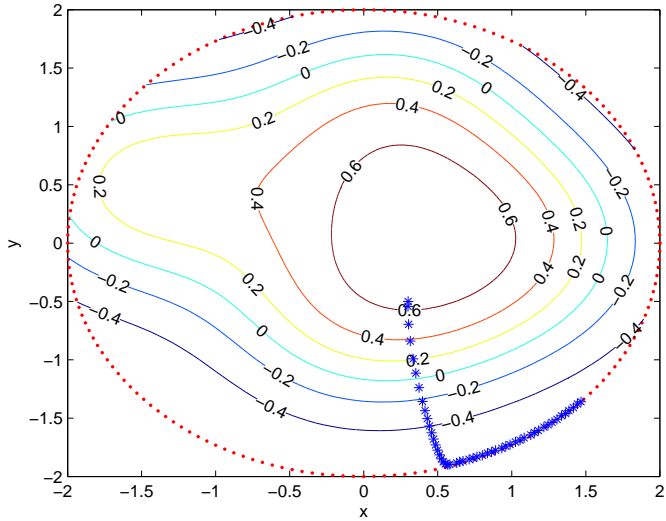, width=4.0in}}
\caption{\label{fig:singleagentflat} 
Top-down view of the surface and trajectory in Fig.1. Level sets of the
projected surface are shown. The boundary of the surface is projected onto the
circle of radius 2.}
\end{figure}

We solve the Eq. (~\ref{rateequations} ) subject to
Eq. ~\ref{equation:Constraint}  by using Runge-Kutta method in time
to obtain the trajectory shown in Fig. ~\ref{fig:singleagent}, where the
initial conditions were $x_0=0.3$ and $y_0=0.5$.
We also show a top-down view of the same trajectory in
Fig. ~\ref{fig:singleagentflat} where the level set of the surface are shown
as well as the boundary of the surface (the circle of radius two) on which the
agent moves.

To achieve movement along the boundary we calculate $C_0$ 
using Eq. (~\ref{confunction} ) for $x$ and $y$ corresponding to coordinates on the circle 
of radius 2 where the surface is defined (any other curve in $x-y$ space could
have been considered).

\begin{figure}
\centerline{\epsfig{figure=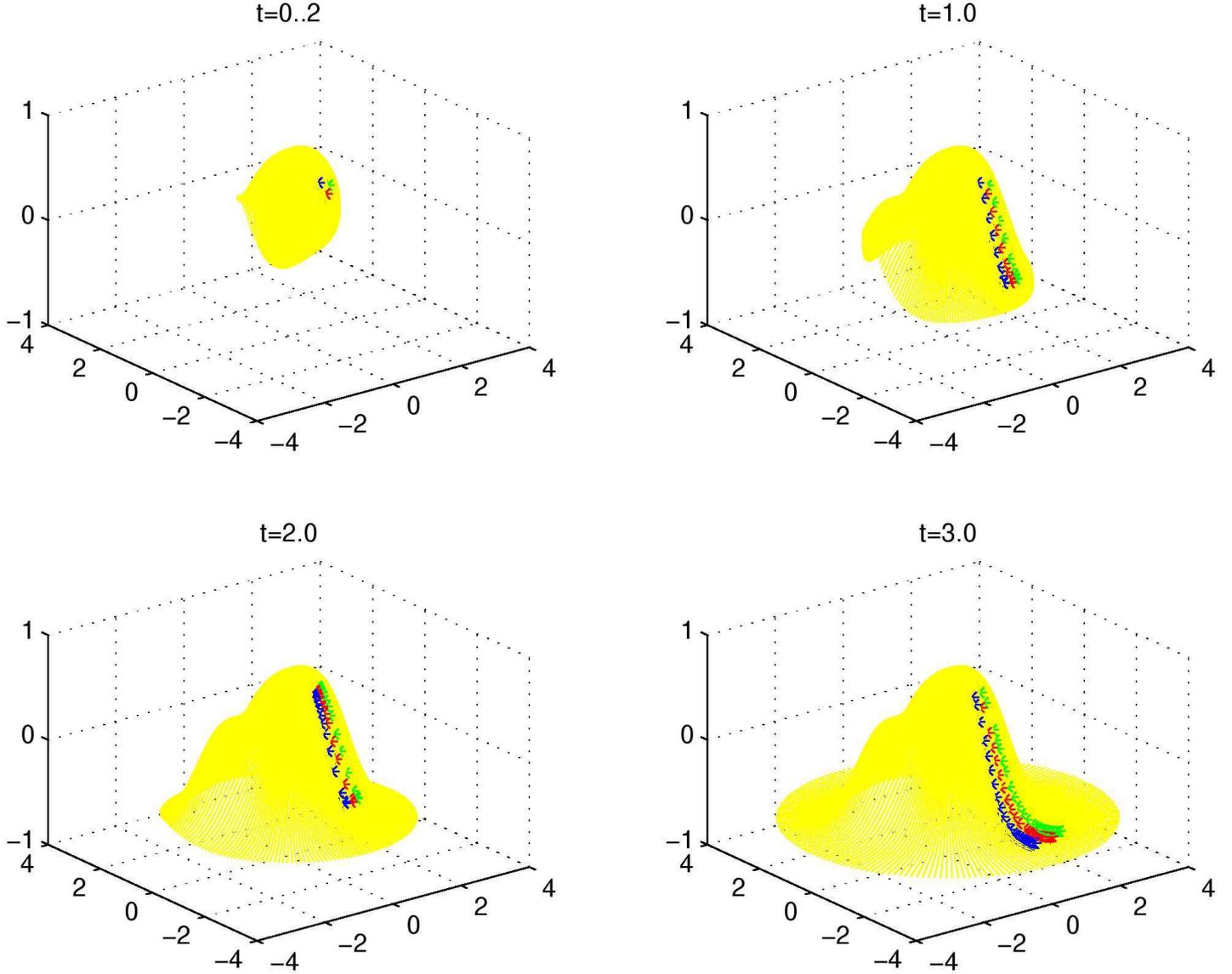, width=7.5in}}
\caption{\label{fig:tripletrajectory} Three nearby trajectories
evolving on a concentration surface whose boundary expands in time at a
constant speed. The four figures show the trajectories at $t=0.2, 1.0,
2.0 ,3.0$. Solid lines denote the evolving concentration surface.}
\end{figure}

The algorithm can be further extended to allow a vehicle to target a boundary
which moves in time. In this case,  Eq. (~\ref{rateequations} ) is used again, the only
difference being that $C_0$ is now a function that varies in time as well in
space. 
In Fig. ~\ref{fig:tripletrajectory} we show three nearby trajectories following a boundary which moves
in time at a constant speed $v_0=2$.  Four different instances in time are
shown at $t=0.6, 1, 1.5$ and $t=2.0$.

\begin{figure}
\centerline{\epsfig{figure=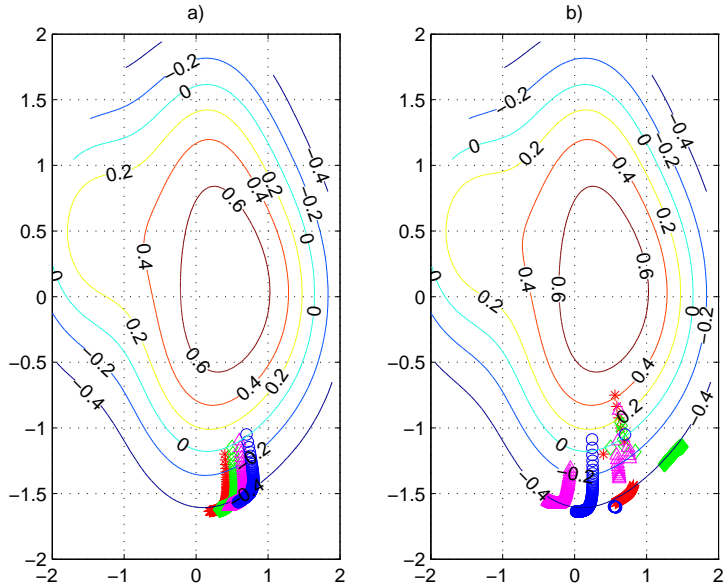, width=4.5in}}
\caption{\label{fig:sparsing} 
Four trajectories, on the concentration surface starting with 
the same initial conditions. (a) There is no communication between agents. 
The trajectories evolve in time independently. 
(b)Similar to (a) but with a sparsing term added into the equations 
of motion, so that agents that get too close repel each other.}
\end{figure}

The existence and uniqueness theorem for ordinary differential
equations \cite{CodLev} guarantees that the trajectories
of this motion will not intersect. In realistic situations when agents are used,
collision among agents can become a concern. For such a situation we add
a sparsing term in Eq.  ~\ref{rateequations} meant to repel nearby
vehicles \cite{MEBS}. This term varies the velocity of a vehicle
according to the following equations:

\begin{equation}
\frac {dv_i}{dt} = \sum _{{v_j}_{j \ne i}} \nabla U(v_j,v_i), \qquad
v_i=(x_i,y_i) \label{spars}
\end{equation}
where the repulsion kernel is given by
$U(v,w)=C_rexp(-\frac{|v-w|}{l_r})$. Here the parameter $l_r$ is a
length scale. 
In what follows $C_r=10$, and $l_r=0.5$.
This term is set non-zero only when the distance between vehicles 
is less than a given cut-off distance $R$; i.e., a maximum distance at which the
sparsing is active.
In Fig. ~\ref{fig:sparsing}b) we show 4 trajectories whose motion is given by the 
Eq. ~\ref{rateequations} to which the sparsing term in Eq. ~\ref{spars} has
been added. For comparison the same trajectories without the sparsing term are
shown in Fig. ~\ref{fig:sparsing}a). When sparsing is added, we see the 4 trajectories,
shown in different colors, intermingle several times on the surface
until smooth trajectories are established. This differs from the orderly
trajectories that can be seen in
Fig. ~\ref{fig:tripletrajectory}.

The above algorithm is used when only a  few independent vehicles are available. When a sufficient
number of vehicles are available, direct interaction among nearby vehicles can be
introduced. This accounts for communication among vehicles in real situations
and for creating desirable patterns of the motion of a swarm, since a swarm
motion is essentially defined as creating various overall patterns and behaviors out of
local interaction rules.
In what follows the approach will be to view the group of agents as a continuous curve
moving and expanding in time. A nice continuum  mechanical analogy
occurs when  the motion is being thought of as  an 
elastic band that expands and bends aiming to
fit around a given boundary \cite{KWT}. This approach will be shown in the next section.

\section{Modeling the swarm behavior as an energy minimizing curve}

The algorithm we develop is meant to be used by vehicles with sensing
capabilities moving through a medium of variable concentration and aimed 
at detecting a certain object such as the boundary of a surface.
Interaction between vehicles can be conceived in various ways leading to a
variety of behaviors of the overall swarm. In what follows we adopt the point
of view of considering the group of agents modeled  as an elastic object 
with elastic and bending forces acting between nearby agents \cite{KWT},\cite{MB}. 
Achieving the desired goal amounts to solving a minimization problem.
We summarize the related theory of this approach in this section and give 
the corresponding numerical discretization in the next section along with 
numerical examples.

The collective motion is modeled as the motion of a deformable contour
which expands and wraps around the boundary, then moves around the boundary. 
Consequently it will be represented by a mapping:

$$\Omega=S^1\rightarrow R^2, s\mapsto v(s)=(x(s),y(s))$$
where $S^1$ is the periodic unit interval. Periodic boundary conditions are
assumed since any agents in the swarm are considered to be identical. 
The algorithm consists in minimizing a certain functional $E$ over a space of
admissible deformations $F$. This functional $E:F\rightarrow R$ represents 
the energy of the contour and has the following form:

$$E(v)=\int_{\Omega}[w_1|v_s(s)|^2+w_2|v_{ss}(s)|^2+P(v(s))]ds\label{Eq:energy1},$$
where the subscripts denote differentiation with respect to the Lagrangian
parameter s which is the arc length, and P is a function of the environmental data. 
P also is designed to incorporate the goal dynamics we are
interested in; i.e. the boundary of a surface. 
The other two terms of the functional $E$ account for the mechanical properties
of the contour of agents; i.e.,  elasticity and bending.
The first  term makes the contour have stretch and the second
 term makes the contour bend \cite{KWT}.
The choice of the parameters $w_1$ and $w_2$ determine the elasticity and the
rigidity of the contour.

If $v$ is a local extremum of $E$ it satisfies the associated Euler-Lagrange
equation:

\begin{equation}
\partial_s (w_1v_s)-\partial_{ss}(w_2v_{ss})+\nabla P=0\label{euler-lagrange}
\end{equation}
with periodic boundary conditions. 
As before
let $P=(C-C_0)^2$.
The relationship between potential
functions and desired goals is discussed comprehensively in \cite{Sapiro} and
 \cite{KWT}.

Since the collective motion we are designing takes place in time,
we approximate the solution of Eq. (~\ref{euler-lagrange}) as the steady state
of the following partial differential equation for the contour $v$:

\begin{equation}
\partial_{t} v=\partial_s (w_1v_s)-\partial_{ss}(w_2v_{ss})+\nabla P,
\label{contour}
\end{equation}
in which the right hand side is the Euler-Lagrange Eq. ~\ref{euler-lagrange}.
This equation describes how each point on the active contour $v$ 
should move in order to minimize the functional $E$.
In the next section we show how this equation can be discretized and used in 
numerical examples. 

 \section{Collective Motion Algorithm along a Virtual Contour}

In this section we present an algorithm for collective motion based on the
minimization approach presented above. 
This algorithm is obtained  discretizing Eq. (~\ref{contour}) 
and considering each spatial discretization point
as representing an agent. Thus a group of agents will evolve according to a
time-dependent scheme and will move along a virtual contour which is an
approximation of the curve $v$ solving Eq. (~\ref{contour}).

We rewrite Eq. (~\ref{contour}) as follows:
\begin{eqnarray}
\partial_{t} x&=\alpha x_{ss}-\beta x_{ssss}+\partial_x P \nonumber \\
\label{eqxy}\\
\partial_{t} y&=\alpha y_{ss}-\beta y_{ssss}+\partial_y P \nonumber \\
\nonumber
\end{eqnarray}
where as previously $P(x,y)=| (C(x,y)-C_{0})|^2$.
Using finite-differencing to discretize the derivatives with respect to the
Lagrangian parameter $s$, we obtain:
\begin{eqnarray}
 \partial_{t} x_i = & \frac{\alpha}{h^2} (x_{i+1}-2x_i+x_{i-1})-\frac{\beta}{h^4} 
(x_{i-2}-4x_{i-1}+6x_i-4x_{i+1}+x_{i+2}) \nonumber \\
  &+\partial_x P(x_i,y_i)& \nonumber \\
\label{eqdiscrete}\\
\partial_{t} y_i = &\frac{\alpha}{h^2} (y_{i+1}-2y_i+y_{i-1})-\frac{\beta}{h^4} 
(y_{i-2}-4y_{i-1}+6y_i-4y_{i+1}+y_{i+2}) \nonumber \\
 &+\partial_y P(x_i,y_i)& \nonumber \\
\nonumber 
\end{eqnarray}

This discretization gives the equation of motion for a group of agents occupying
the positions $(x_i,y_i)$. We define thus the motion of the mobile agents by a
set of coupled ODEs for the positions of the $N$ agents. In what follows we
take $\alpha=0.01$ and $\beta=0.0001$ as in \cite{BKM}.

The equations of motion  given by (~\ref{eqdiscrete}) will evolve agents
towards the boundary of the concentration surface. Once on the
boundary, movement along the boundary is generated by  adding a term parallel
to the normal to the surface as in (~\ref{rateequations}).
This results in the following scheme:

\begin{eqnarray}
\partial_{t} x_i=&\frac{\alpha}{h^2} (x_{i+1}-2x_i+x_{i-1})-\frac{\beta}{h^4} 
(x_{i-2}-4x_{i-1}+6x_i-4x_{i+1}+x_{i+2})\nonumber \\
&+\partial_x P(x_i,y_i)-
\omega\frac{\partial_yC(x_i,y_i)}{|\nabla C(x_i,y_i))|} +
\sum _{j \ne i} {\partial _x U}((x_i,y_i),(x_j,y_j))\nonumber \\
\label{eqdisc2}\\
\partial_{t} y_i=&\frac{\alpha}{h^2} (y_{i+1}-2y_i+y_{i-1})-\frac{\beta}{h^4} 
(y_{i-2}-4y_{i-1}+6y_i-4y_{i+1}+y_{i+2})\nonumber \\
&+\partial_y P(x_i,y_i)+
\omega\frac{\partial_x C(x_i,y_i)}{|\nabla
  C(x_i,y_i))|}+\sum _{j \ne i}{\partial _{y} U}((x_i,y_i),(x_j,y_j)).\nonumber \\ 
\nonumber
\end{eqnarray}
The last term represents the sparsing term that has been added as in
Eq. ~\ref {spars}. In the examples below we let $\omega = 2.0$ and the time derivative 
in Eq.  ~\ref{eqxy} has been discretized using the
forward Euler method. With this time
discretization it follows that one and respectively two-nearest neighbor 
communication will occur, as seen from the discretization
of the diffusion term and the discretization of the 4th order derivative.
Stability issues as well an implicit version concerning this algorithm 
have been discussed in \cite{MB} where it is shown that stability of the
forward Euler scheme requires the condition ${\Delta
  t}<\frac{{h}^4}{8\beta+2\alpha h^2} $, where $h$ stands for the distance between nearby agents.

\begin{figure}
\centerline{\epsfig{figure=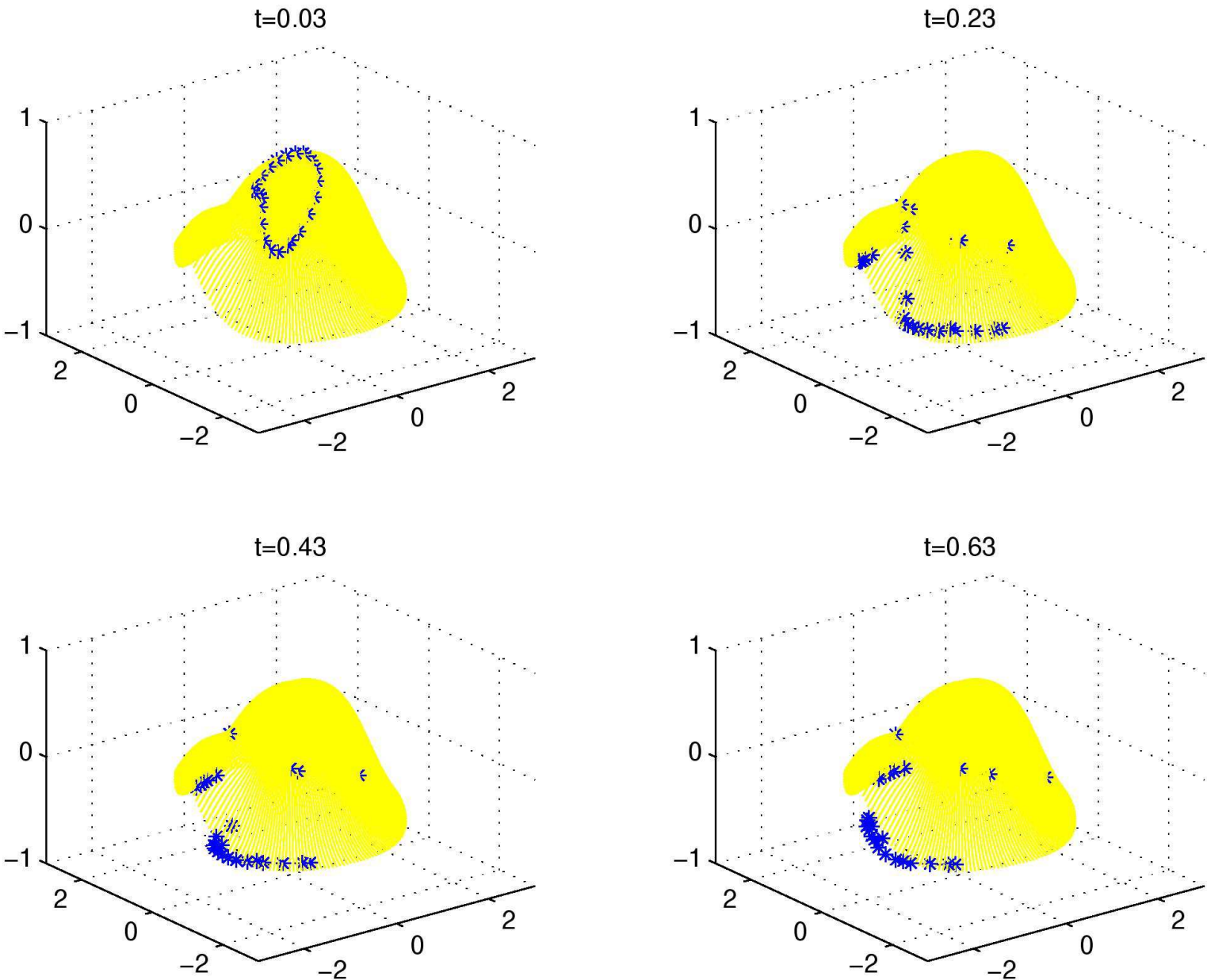, width=8.0in}}
\caption{\label{fig:25swarm} The picture illustrates
a swarm of 25 vehicles at different instances in time. The algorithm 
accurately follows a boundary variable in space.
The motion is shown at $t=0.03$ then at $t=0.23,0.43$ and $t=0.63$.}
\end{figure}

\begin{figure}
\centerline{\epsfig{figure=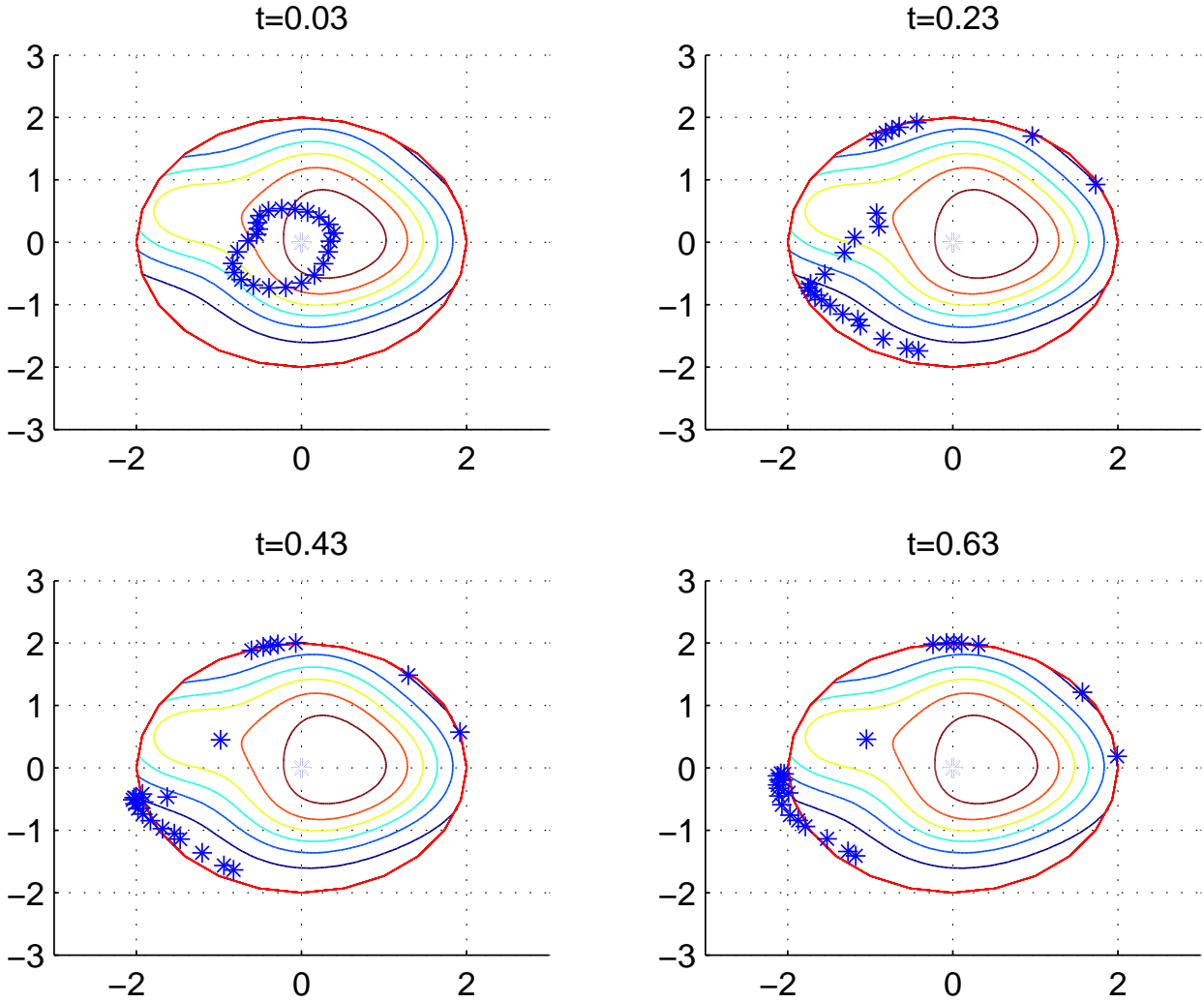, width=8.0in}}
\caption{\label{fig:25swarmflat} The picture gives a top-down view
of the swarm in Fig.5. The algorithm accurately follows a boundary variable 
in space which, when projected forms a circle of radius 2.}
\end{figure}

Using Eq. (~\ref{eqdisc2}) one can effectively target and track a level
curve of a concentration surface. We have allowed in Eq. (~\ref{eqdisc2}) the concentration on the
boundary to be a function of $x$ and $y$. In Fig. ~\ref{fig:25swarm}, we show a swarm of 25
vehicles at four different instances in time, each one 20 time steps apart with
a time step $\Delta t=0.01$. We show the swarm at the
beginning of the motion which starts in the center of the surface, then we
show three later pictures of the same swarm as it circles the surface
boundary. Fig. ~\ref{fig:25swarmflat} is a top-down view of
Fig. ~\ref{fig:25swarm} in which the projection of the surface boundary, (i.e.
the circle of radius two) is marked so that one can clearly see the movement
of the agents along this boundary. Figure ~\ref{fig:25swarmflat} shows a top-down view of
Fig. ~\ref{fig:25swarm} in which the projected boundary of the surface (that
is the circle of radius two) is marked and this enables one to see the
accuracy of the motion of the swarmers on this boundary.

\begin{figure}
\centerline{\epsfig{figure=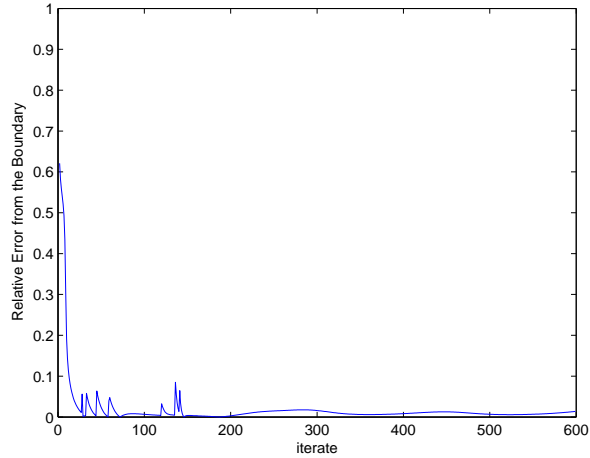, width=3.5in}}
\caption{\label{fig:pointerror} The picture shows
the relative error corresponding to one of the trajectories in Fig. ~\ref{fig:25swarm},
which is seen to stay below $2 \%$}
\end{figure}

In Fig. ~\ref{fig:pointerror} we pick one of the vehicles shown in
Fig. ~\ref{fig:25swarm}
and show the relative error evolution for this particular trajectory. 
The chosen trajectory is the one that starts at
$x_0=0.4 cos(\frac{3\pi}{4})\nonumber$,
$y_0=0.4 sin(\frac{3\pi}{4})\nonumber$,
and it shows a relative error which is below $2 \%$. The rest of the vehicles
behave similarly. 

\begin{figure}
\centerline{\epsfig{figure=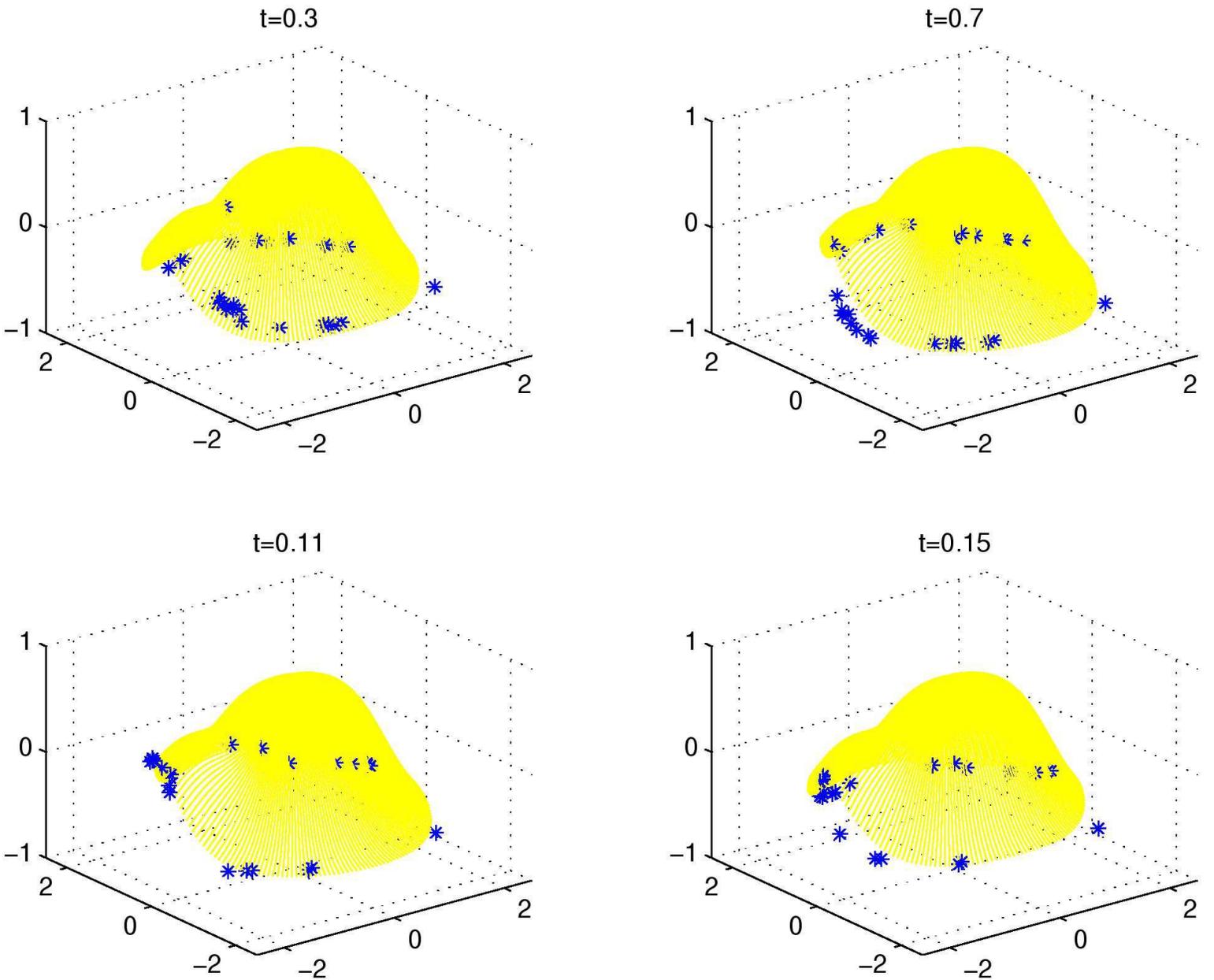, width=8.0in}}
\caption{\label{fig:bdrysimpletime} The picture shows four instances in
 time of the motion of a swarm of vehicles on a time-varying surface.
The motion is shown at $t=0.3, 0.7, 0.11, 0.15$, the time step was $\Delta
 t=0.01$}
\label{timeswarm}
\end{figure}

\begin{figure}
\centerline{\epsfig{figure=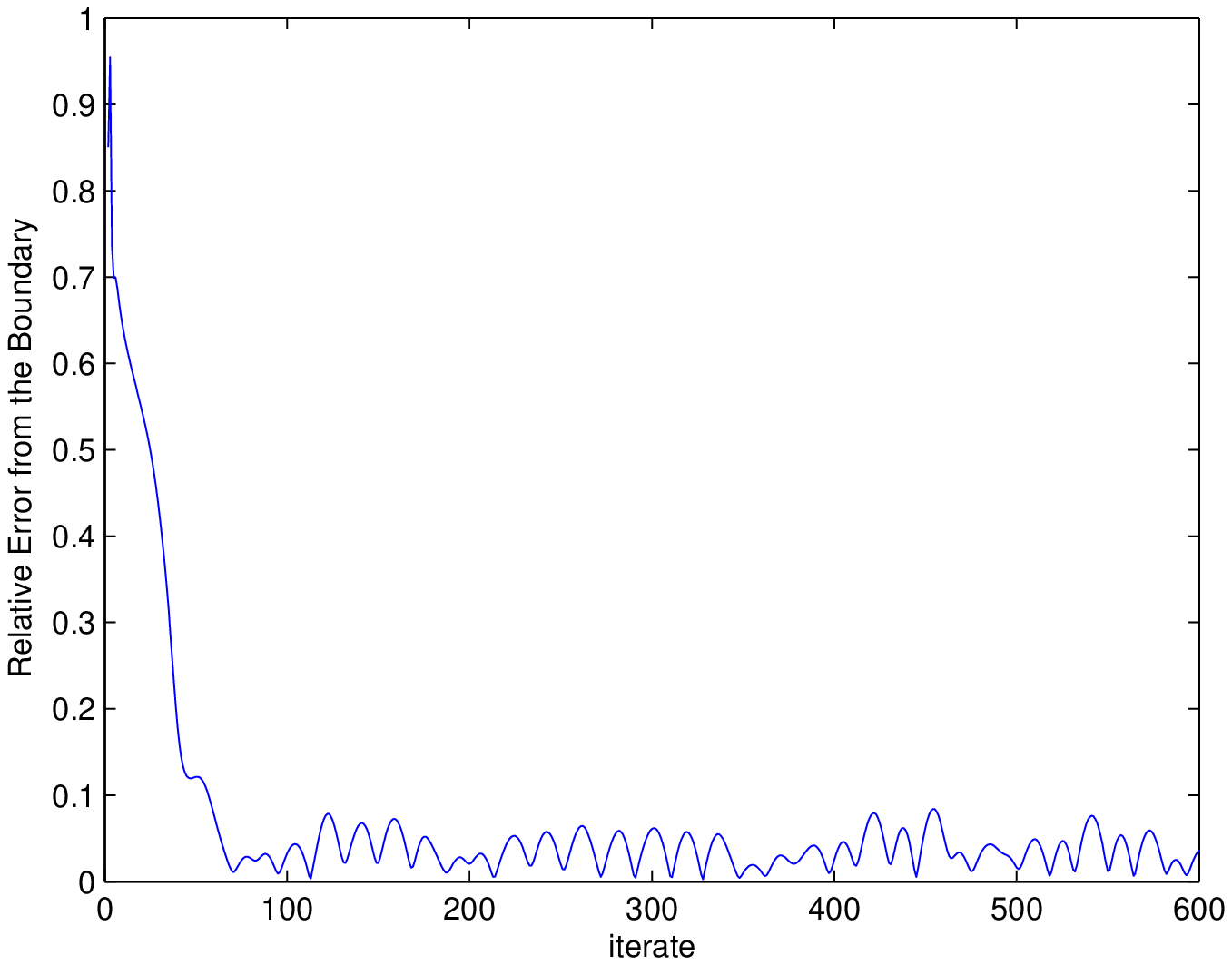, width=4.5in}}
\caption{\label{fig:timeerror2} The relative error in the motion
of a single vehicle chosen at random out of the 25 trajectories in Fig.
~\ref{timeswarm}
The error averages $3.6 \%$.}
\end{figure}

Developing the algorithm further, in Eq.(~\ref{eqdisc2}), we allow 
both the concentration $C$ and the boundary $C_0$
to be functions of time as well as space. 
To simulate variation in time we consider the same
surface as before given by (~\ref{confunction}) and vary both $x$ and $y$
periodically in time with different frequencies
according to: 
$$x=A_0cos(\frac{2\pi t}{40}) \qquad 
y=A_0sin(\frac{2\pi t}{30})\nonumber.$$
This results in a generic, time-dependent change of the whole surface.
Here we show in Fig. \ref{fig:bdrysimpletime} four images of the motion at
various moments in time.  
In Fig. ~\ref{fig:timeerror2} we show the relative error associated with the motion of one of the 
trajectories in Fig. ~\ref{fig:bdrysimpletime}. We chose again the same initial conditions
as for Fig. ~\ref{fig:pointerror}.
This error averages out to $3.6 \% $ when $A_0=0.1$ and stays below $5 \%$.
If we raise $A_0=0.15$, the corresponding average error will be $5.3 \%$.

\section{Collective motion in 3D}

The above algorithm can be realized in 3D by a similar variational
formulation. Initially the minimization problem was introduced in image 
processing \cite{ICLCNA} as a deformable surface model meant to evolve to a certain
location and shape. We extend that idea to the motion of a collective
of agents. We  consider the group of agents placed on
a deformable surface which evolves in time to fit a certain boundary surface.
The deformation assumes elastic and bending forces that are internal to the
deformable surface. The forces have to balance the so-called external forces
that are stretching this surface so that it will fit a boundary surface or
some other type of dynamical object that the moving swarms have to reach. 
We adopt, as in \cite{ICLCNA}, the model consisting of minimizing an energy functional $E$
over a set of parameterized admissible surfaces $v$ defined by the mapping:

\begin{eqnarray}
\Omega =&[0,1]\times [0,1] \rightarrow R^3\nonumber\\
\label{3Dv}\\
v(s,r)=&(x(s,r),y(s,r),z(s,r)).\nonumber\\
\nonumber
\end{eqnarray}
The associated energy $E$ is given by:

\begin{eqnarray}
E(v)=&\int_{\Omega} [w_{10}|v_s(s)|^2 + w_{01}|v_{r}(s)|^2 + 2w_{11}|v_{sr}(s)|^2 \nonumber \\
&
+ w_{20}|v_{ss}(s)|^2 + w_{02}|v_{rr}(s)|^2 + P(v(s,r))] ds dr 
\label{energy}
\end{eqnarray}
where $P$ is associated to the task to be achieved by the algorithm. 
The boundary is now a 2D surface in space.
Similar to the 2D case we let $P(x,y,z)=(C(x,y,z)-C_{0})^2$ where now $C$ and $C_0$ 
are functions of $x$,$y$ and $z$. 
The other terms in Eq. ~\ref{energy} determine how the motion in space of the
surface $v$ will take place. Continuing our analogy from continuum
mechanics, these terms model internal forces that 
are acting on the shape of the surface $v$ as it moves towards its goal.
Consequently the shape of the surface $v$ will depend on the elasticity
coefficients $w_{10},w_{01}$, the rigidity coefficients $w_{20},w_{02}$
and the resistance to twist $w_{11}$. We also constrain the surface $v$ by assuming
periodic boundary conditions. 

A local minimum $v$ of $E$ satisfies the associated Euler-Lagrange equation:

\begin{eqnarray}
&-w_{10}v_{ss}(s,r) - w_{01}v_{rr}(s,r) + 2w_{11}v_{ssrr}(s,r) 
 +w_{20}v_{ssss}(s,r) \nonumber \\
 &+ w_{02}v_{rrrr}(s,r)  = - \nabla  P(v(s,r))
\label{ELspace}
\end{eqnarray}
to which boundary conditions are added. 
Equation  ~\ref{ELspace} represents the necessary condition for a minimum of $E$, ($E^{'}(v)=0$).
A solution of  Eq. ~\ref{ELspace} can be seen as either realizing the equilibrium
between internal and external forces or as reaching a minimum of the energy $E$.
Since there may be many local minima,  appropriate initial data $v_0(s,r)$,
leading to the desired minimum 
is assumed when solving the associated evolution equation, in which we add a temporal 
parameter $t$:

\begin{eqnarray}
&\frac {\partial v}{\partial t}-w_{10}v_{ss}(s) - w_{01}v_{rr}(s) +
  2w_{11}v_{ssrr}(s) 
+ w_{20}v_{ssss}(s) \nonumber \\
&+ w_{02}v_{rrrr}(s) = - \nabla  P(v(s,r)).
\label{ELtemp}
\end{eqnarray}
A solution to the static minimization problem is obtained when the solution
$v$ converges as $t$ tends to infinity, which means that the term
$\frac {\partial v}{\partial t}$ vanishes leading to a solution of the static
problem. 

The evolution Eq. (~\ref{ELtemp}) will be used to model the motion of a swarm in
3D. We discretize this equation in space using finite-differences and each
discrete spatial point will stand for the position of a vehicle
proceeding analogously to the 2D case in the previous section. 
Thus this swarm will be designed as lying on the surface $v(s,r)$ and
moving towards the assigned boundary with interactions between nearby vehicles
determined by the elasticity and rigidity coefficients in Eq. (~\ref{ELtemp}).
 We apply this discretization (not presented here, but standard and analogous to
the discretization in section IV) to the motion of a specific swarm evolving on a
concentration volume defined in 3D of equation:

\begin{eqnarray}
C(x,y,z) & =  & tanh (F(x,y,z) - \frac {3}{4})\nonumber \\
\label {concfunction}\\
F(x,y,z) & =  & \sum_{i=1}^{4} exp(\frac {-((x-x_i)^2+(y-y_i)^2+(z-z_i)^2)}{\sigma
  ^2})\nonumber \\
\nonumber
\end{eqnarray}
where $(x_i,y_i)=(1,0,-1),
(0,-\frac {1}{2},\frac {1}{2}),(-\frac {3}{2},\frac {1}{2},0),(\frac
{3}{20},1,-\frac {1}{2})$ for $i=1,2,3,4$ and $\sigma =1.0$.

The swarm will move in space through a medium of variable concentration in
space. The swarm will move along the steepest descent direction which is 
given now by a 3-dimensional vector 
$(\frac {\partial P}{\partial x},\frac {\partial P}{\partial y},\frac {\partial P}{\partial z})$.

\begin{figure}
\centerline{\epsfig{figure=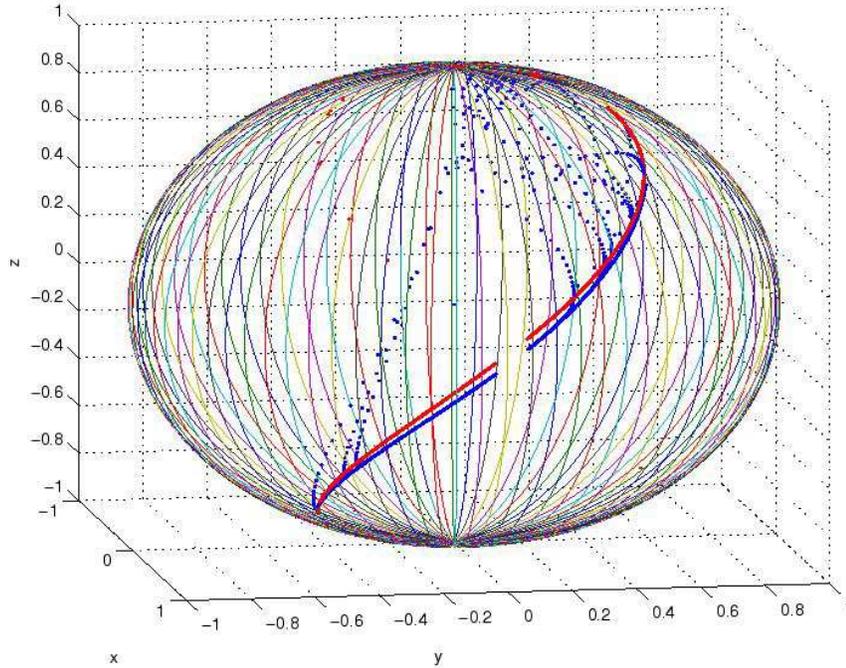, width=4.5in}}
\caption{\label{Fig:desct} 
14 trajectories starting inside the spherical domain, reaching the boundary
and following a specified curve on the boundary.}
\end{figure}

\begin{figure}
\centerline{\epsfig{figure=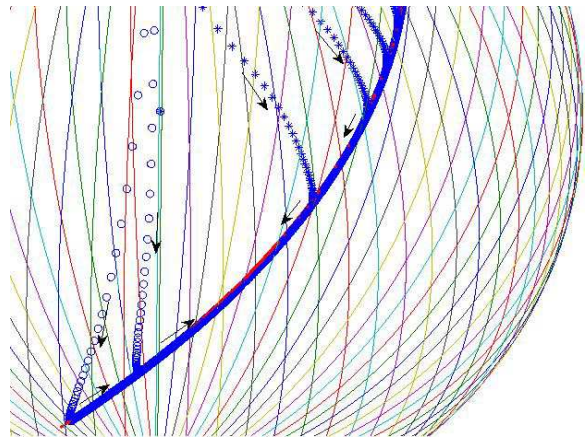, width=3.0in}}
\caption{\label{Fig:desct_bu} 
Blowup of Fig. ~\ref{Fig:desct}, where the arrows show the direction
of the agent's trajectories.}
\end{figure}

A last numerical example shows that the motion of the swarm in 3D can be
directed to reach and track a specific curve on the boundary.
We define a curve in space lying on the spherical boundary. This curve
is defined by the equation:

\begin{eqnarray}
x=cos(\phi)&cos(\phi)\nonumber\\
y=cos(\phi)&sin(\phi)\label{exquisite}\\
z=sin(\phi)&\nonumber\\
\nonumber
\end{eqnarray}
When using the discretization of Eq. ~\ref{ELspace} with this modified 
goal dynamics we obtain the 14 trajectories seen in Fig. 10.
The goal dynamics is introduced in the discretization of Eq. ~\ref{ELspace}
by evaluating $C_0$ at $x$,$y$ and $z$ given by ~\ref{exquisite}.
In Fig. 10 we show both the exact target curve (in red)a
and the curve described by 14 swarmers (in blue)
as they approximate movement along the curve of Eq. ~\ref{exquisite}.
In this figure we remark that the target curve is traveled by the 
swarmers in opposite directions. We also show a detail of  Fig. ~\ref{Fig:desct}
in Fig. ~\ref{Fig:desct_bu} where arrows indicate how the trajectories
evolve from inside the sphere towards the surface boundary and then move along
the prescribed curve.  

As in the 2D case we now vary the concentration in time.
In the concentration equation, Eq. ~\ref{concfunction}, we let x, y and z vary in time
according to the formula:

\begin{equation}
x=A_0 cos(\frac{2\pi t}{40}) \qquad 
y=A_0 cos(\frac{2\pi t}{30})\qquad z=A_0 cos(\frac{2\pi t}{20})
\label{initsp}
\end{equation}
The corresponding figure is not shown here since it is very similar to Fig. 10
with the exception that the sphere undergoes slight deformations in time
due to the motions given in ~\ref{initsp}.
To illustrate the time-dependent case we give in Fig. 12 the relative error
for one agent. Namely the one starting at:
$x_0=0.8cos(\frac{4\pi}{5})cos(8\frac{\pi}{25})$,
$y_0=0.8cos(\frac{4\pi}{5})sin(8\frac{\pi}{25})$,
$z_0=0.8sin(\frac{4\pi}{5})$.
It can be seen in this figure that the relative error is below $5 \%$.

\section{Conclusions}

We have presented an algorithm that allows a formation of agents to move
towards a prescribed boundary and then describe that boundary. The motion
takes place on  stationary or non-stationary surfaces and can be realized as dynamics on a
surface (2D) or as motion in space (3D).

The algorithm represents a new development over snake algorithms used in
image processing \cite{KWT},\cite{BKM}. Those algorithms were primarily
designed to detect level-sets in an image. Here we show that a similar
algorithm will detect boundaries that vary in space and in time.
Our new algorithm, along with the one in \cite{BKM}, directs 
the idea of snake algorithms towards
solving a new problem, namely the problem of controlling motions of swarms. 
Another development was to allow the environment in which the swarm is moving to
depend on time in a quite general manner.

In future work we will extend the current algorithm to detecting curves 
with a more complex topology than the one considered here. 
Our main goal in future developments of these algorithms will be to couple the
equations of motion of the swarm with meaningful dynamical systems with
relevance for biological application, chemical reactions and advective flows.
In this way the abstract goal of reaching a certain surface or feature will be
directed towards  real-life applications.

\begin{figure}
\centerline{\epsfig{figure=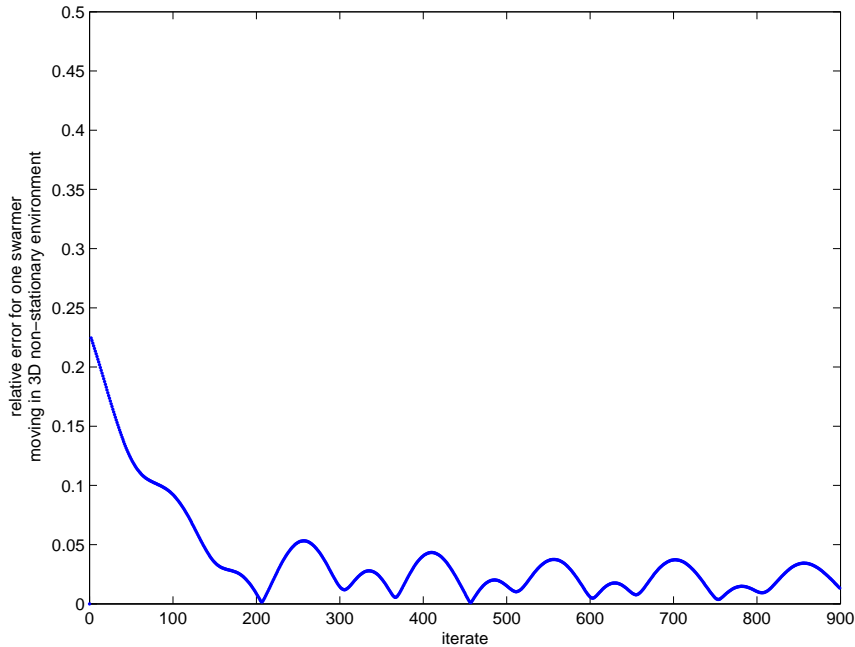, width=4.5in}}
\caption{\label{Fig:error1} The relative error
in the motion of a single swarm reaching and then moving along a space and time-dependent
boundary in 3D.}
\end{figure}

\section{Acknowledgment}

This research has been supported by the Office of Naval Research.

%\bibliographystyle{plain}
%\bibliography{swarm.bib}

\begin{thebibliography}{15}
\bibitem{MB} Daniel Marthaler and Andrea L. Bertozzi, Collective motion algorithms for
determining environmental boundaries, Autonomous Robots, special issue on
Swarming, submitted 2002.
\bibitem{BKM} A.L. Bertozzi M. Kemp, and D. Marthaler, Determining Environmental
Boundaries: Asynchronous communication and physical scales, in Proceedings
of the Block Island Workshop on Cooperative Control 2003, Lecture Notes in
Control and Information Systems, edited by V. Kumar, N. Leonard, S. Morse
(September 2003), Springer-Verlag.
\bibitem{CP} ``Swarming: Network Enabled C4ISR, 13-14 Jan 2003. Conference
Proceedings, http://search.netscape.com/ns/.
\bibitem{BDT} Eric Bonabeau, Marco Dorigo and Guy Theraulaz, ``Swarm
Intelligence- From Natural to Artifical Systems'', a volume in the Santa Fe
Institute Studies in the Sciences of Complexity, Oxford University Press,
1999. 
\bibitem{PEK} J.K.Parrish and L. Edelstein-Keshet, ''Complexity, Pattern 
and Evolutionary Trade-Offs in Animal Aggregation``, 
Science Vol. 284, 5411, pages 99-101, 1999.
\bibitem{CASS} Riccardo Cassinis, ``Landmines Detection Methods Using Swarms of
Simple Robots'', Intelligent Autonomous Systems 6, E.Pagello et al. (Eds)
IOS Press, 2000. 
\bibitem{SD} A. Stentz and M. B. Dias, `` A Free-Market Architecture for
Coordinating Multiple Robots'', CMU-RI-TR-99-42, The Robotics Institute,
Carnegie Mellon University, Dec. 1999. 
\bibitem{BMFST} W.Burgard, M.Moors, D.Fox, R. Simmons and S. Thrun,
``Collaborative Multi-Robot Exploration'', Proceedings of IEEE 
International Conference on Robotics and Automation, April 2000. 
\bibitem{DOK} J.P.Dessai, I.Ostrowski and V.Kumar, ``Controlling formations
of multiple mobile robots'', Proc. of IEEE International Conference on
Robotics and Automation, May 1998.
\bibitem{OFL} Petter Ogren, Edward Fiorelli and Naomi Ehrich Leonard, 
``Formations with a Mission: Stable Coordination of Vehicle Group Maneuvers'', 
Proc. Symposium on Mathematical Theory of Networks and Systems, August 2002.
\bibitem{KWT} Michael Kass, Andrew Witkin and Demetri Terzopoulos,
``Snakes: Active Contour Models'',
International J. of Computer Vision, 321-331 (1988).
\bibitem{Sapiro} Guillermo Sapiro, ``Geometric Partial Differential Equations
and Image Analysis'', Cambridge University Press, 2001. 
\bibitem{CodLev} Earl A. Coddington and Norman Levinson, ``Theory of Ordinary
  Differential Equations'', McGraw-Hill Book Company, 1955. 
\bibitem{MEBS} A.Mogilner, L.Edelstein-Keshet, L.Bent, A. Spiros
Mutual interactions, potentials, and individual distance in a social
aggregation, to be published.  
\bibitem{ICLCNA} Isaac Cohen, Laurent D. Cohen and Nicholas Ayache, ``Using
  Deformable Surfaces to Segment 3D Images and Infer Differential
  Structures'', CVGIP: Image Understanding Vol.56, No.2 September,
  pp. 242-263, 1992. 


\end{thebibliography}

\end{document}